\documentclass[twocolumn,preprintnumbers,amsmath,amssymb,superscriptaddress]{revtex4}

\usepackage{hyperref}%
\usepackage{graphicx}%
\usepackage{dcolumn}%
\usepackage{bm}%

\def\lsim{\mathrel{\raise.3ex\hbox{$<$\kern-.75em\lower1ex\hbox{$qf$}}}}
\def\gsim{\mathrel{\raise.3ex\hbox{$>$\kern-.75em\lower1ex\hbox{$\sim$}}}}

\begin{document}

\title{Comments on ``First Results of the Phase II SIMPLE Dark Matter Search''}

\author{C.E. Dahl}
\affiliation{Enrico Fermi Institute, KICP and Department of Physics, University of Chicago, Chicago, USA}

\author{J. Hall}
\affiliation{Fermi National Accelerator Laboratory, Batavia, USA}

\author{W.H. Lippincott}
\affiliation{Fermi National Accelerator Laboratory, Batavia, USA}

\date{\today}


\maketitle

The SIMPLE Collaboration has reported results from their superheated C$_2$ClF$_5$ droplet 
detectors~\cite{SIMPLE1,SIMPLE2}, including a description 
of acoustic discrimination between $\alpha$ decays and nuclear recoils.  Our concern is that the events in the neutron calibration data and the events identified as neutrons in the physics data are not drawn from the same parent distribution. This fact calls into question the identification of the background events as neutrons, the use of the calibration data to define the acceptance of WIMP-induced nuclear recoils, and the observation of discrimination against $\alpha$'s. 

\begin{figure}[hb]
\includegraphics[width=260 pt]{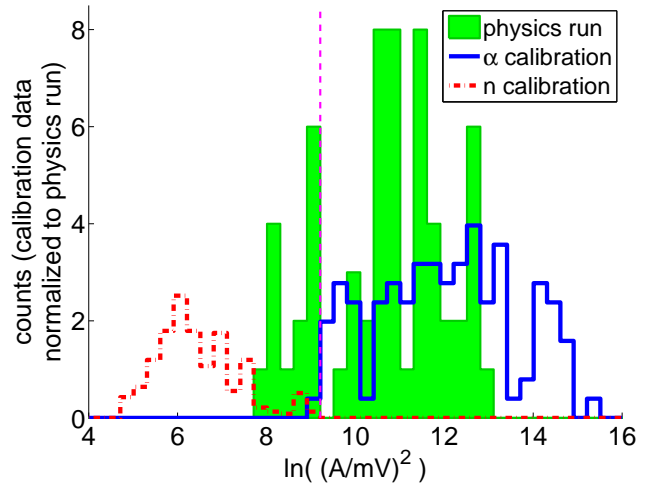}
\caption{\label{fig:SIMPLEdata} 
Histogram of the data in Fig.~1 of~\cite{SIMPLE1} projected on its y-axis. The physics data 
to the left of the dashed line are identified as neutrons and those to the right as $\alpha$'s.  The y-axis is in counts for physics 
data, and each calibration histogram is scaled to have the same area as the corresponding region of the physics 
data. If the calibrations are representative, the physics data should match the sum of the calibration histograms.}
\end{figure}

Figure~1 of~\cite{SIMPLE1} (reproduced in Fig.~\ref{fig:SIMPLEdata} below as a 
projection of the original figure onto its y-axis) shows four 
classes of data points representing three sets of data:  
$\alpha$-spiked data, neutron calibration data, and the physics data. %

The authors draw a line between the acoustic distributions observed in the calibration runs, defining events 
 with amplitude below this line to be neutrons and those with 
amplitude above to be $\alpha$'s.

The fundamental issue with the SIMPLE analysis is that the neutron 
calibration data set and the events identified as neutrons in the physics 
data do not have the same parent acoustic distribution, as shown in our Fig.~\ref{fig:SIMPLEdata}. In the neutron calibration data, $89\%$ of the events have ln((A/mV)$^2) < 7.5$, while none of the 14 identified neutrons from the physics data have  ln((A/mV)$^2) < 7.5$. 
A two-sample K--S test gives a probability $<10^{-10}$ that the datasets are 
samples of the same parent distribution.
This raises two immediate questions:  are the physics events identified as neutrons 
truly neutrons?  If so, why is their acoustic distribution different from the calibration distribution?

The fact that the calibration and physics distributions do not agree directly 
affects the determined WIMP sensitivity.  
The neutron calibration source data is used to calculate an acceptance of $97\%$ 
for the acoustic cut for nuclear recoils, setting the sensitivity of the 
physics modules to dark matter. As the data show that the acoustic distribution 
of the physics modules does not match the calibration, we believe that the actual 
acceptance of the acoustic cut when applied to the physics data is unknown.
 
We note that neutron sources were applied to neither the physics nor the $\alpha$-spiked modules. Given the mismatch between the neutron calibrations and physics data, we wonder
what neutrons would sound like in the $\alpha$-spiked modules.
The authors' claim of discrimination relies on the assumption that the 
acoustic distributions in the $\alpha$-spiked and neutron-calibration modules would agree.

If acoustic discrimination is indeed the source of the difference between the two calibration datasets,
 then $\alpha$ decays 
produce 400 times the acoustic power of nuclear recoils in the SIMPLE detectors. 
This is in tension with results from PICASSO and COUPP, using C$_4$F$_{10}$ 
superheated droplet detectors and CF$_3$I bubble chambers, where $\alpha$ 
decays produce only 4 times the acoustic power of nuclear recoils~\cite{PICASSOdisc,COUPPnumi}.

The SIMPLE Collaboration's interpretation of their data is not supported by the calibrations 
presented in the Letter. As a result, we must call into question the observation of acoustic 
discrimination and the reported dark matter sensitivity.  
We look forward to the resolution of these concerns
 as more information becomes available. 

\end{document}